**Title:** Steering visible Dyakonov surface waves at the hyperbolic metasurface

**Authors:** Yan Li, Jingbo Sun*, Yongzheng Wen, Ji Zhou*

**Affiliations:**

State Key Laboratory of New Ceramics and Fine Processing, School of Materials Science and Engineering, Tsinghua University, Beijing 100084, China

Correspondence and requests for materials should be addressed to:

jingbosun@tsinghua.edu.cn

zhouji@tsinghua.edu.cn

**ABSTRACT**

Dyakonov surface wave existing at the interface with anisotropy offers a promising way of guiding light in two-dimension with almost no loss. However, predicted decades ago, the experimental demonstration of the Dyakonov surface wave seems always challenging for the weak anisotropic indices from the natural materials. Here we experimentally demonstrated a Dyakonov surface wave mode propagating in a hyperbolic metasurface at the visible frequency. Dyakonov surface waves at the two surfaces of the metasurface can be supported simultaneously by the hyperbolic anisotropy and form a Dyakonov typed mode with low loss and a large allowed angle band. A detailed theoretical analysis and numerical simulations prove that the electric field of such a surface wave mode shows transverse spin, whose direction is determined by the orientations of the hyperbolic anisotropy and surface normal, based on which we experimentally observed the photonic spin Hall effect of the surface wave mode in our metasurface.

# I. INTRODUCTION

Optical surface waves are tightly confined light at the nanoscale, which play a significant role in the minimization and dimension reduction of photonic circuits [1- 5]. So far, most achievements in these areas have been accomplished using surface plasmon polaritons at the interface between a metal and a dielectric medium [6- 15], which have produced remarkable progress, such as plasmon based source [6, 7], guiding [8- 10], routing [11, 12], and modulations [13- 15] in the nano photonics field in recent years. However, all surface plasmon polaritons suffer from high propagation loss due to their transmission on metal, especially in visible range [16].

Dyakonov surface waves (DSWs) were predicted to solve this problem since they can be supported at the interface formed by two transparent materials while with at least one of them to be anisotropic [17]. As an instance, at an interface between an isotropic (with refractive index $n_c$) and a uniaxial anisotropic (with anisotropic refractive indices $n_o$ and $n_e$) media, when the index relation fulfils the condition: $n_o < n_c < n_e$, DSW can be generated within an angle band $\Delta\theta$ with respect to the optical axis of the anisotropic material. According to these requirements, it turns out that DSW is always hard to observe due to the limited anisotropy of natural crystals, from which only few materials are able to match the index condition but with extremely tiny $\Delta\theta$ [18- 25]. One intelligent approach is to add an isotropic nano-sheet in-between the isotropic and anisotropic media, where guided modes enabled by DSW are formed and exhibit excellent long distance propagation behavior in the ultra-thin film as well as an increased allowed angle band [26]. However, this method involves three different materials which satisfy certain conditions regarding their refractive indices, and always requires an index matching liquid. Meanwhile, hyperbolic metamaterials with extremely strong anisotropy have been considered to support the propagation of DSWs [27- 34]. But the fabrication of a bulky hyperbolic metamaterial with in-plane anisotropy is rather challenging, especially in the visible range [35]. Here, we explore DSWs propagation using an ultra-thin hyperbolic film on a quartz glass substrate, and find that the hyperbolic anisotropy can simultaneously compliment both the air and glass as isotropic counterparts, thus it can support DSWs on its two interfaces. Together, these form one propagation mode of unique properties, such as a much larger allowed angle band, low loss and transverse spin, based on which we observe the photonic spin Hall effect (PSHE) experimentally.

## II. Theory

We begin by studying the unique properties of the DSW mode in a hyperbolic metasurface. By utilizing the hyperbolic material with extraordinary permittivity $\varepsilon_e$ and ordinary permittivity $\varepsilon_o$ ($\varepsilon_e > 0 > \varepsilon_o$) as the anisotropic counterpart of either air (with permittivity of $\varepsilon_{air}$) or a quartz glass substrate (with permittivity of $\varepsilon_{glass}$), the strict permittivity condition in a conventional DSW ($\varepsilon_e > \varepsilon_c > \varepsilon_o > 0$) system can be alleviated to be $|\varepsilon_o| > \varepsilon_{air}, \varepsilon_{glass}$ [21]. This can be naturally satisfied at the interfaces between air/ hyperbolic metamaterial and hyperbolic metamaterial/ glass. By decreasing the thickness of the hyperbolic metamaterial to the subwavelength level to form a metasurface, both surface waves get coupled together and form a surface mode in the hyperbolic metasurface, as shown in Fig. 1(a). If we assume the wave vector $\boldsymbol{k} = k_0 \boldsymbol{N}$ of the DSW mode to be along $x$-axis, and the optical axis (OA) of hyperbolic metasurface is oriented to an angle of $\theta$ with respect to $x$-axis, fields in the isotropic materials are in a hybrid polarization. This polarization is a linear combination of the transverse electric (TE) and transverse magnetic (TM) modes [17]:

$$\text{TE: } E_{TE} = A_{TE}(0 \quad 1 \quad 0)\exp(-k_0\beta_c z)\exp\left[i(k_0 N x - \omega t)\right] \tag{1a}$$

$$\text{TM: } E_{TM} = A_{TM}(-i\beta_c \quad 0 \quad N)\exp(-k_0\beta_c z)\exp\left[i(k_0 N x - \omega t)\right] \tag{1b}$$

where $k_0$ is the wave vector in vacuum, $N$ is the effective refractive index of the DSW modes and $\beta_c = \sqrt{N^2 - n_c^2}$ is the decay constant along z direction in the isotropic material. $A_{\text{TE}}$ and $A_{\text{TM}}$ are the amplitudes of the TE and TM mode, respectively. Here, $n_c$ is either the refractive index of the air or the glass substrate. According to the boundary condition [17], fields at the top interface can also be expressed via equation (1) when $z = 0$. The imaginary factor $i$ for the longitudinal component $E_x$ shows $\pi/2$ out of phase with respect to the transverse component $E_r = E_y + E_z$, enabling a transversely spinning electric field, as shown in Fig. 1(b). According to the numerical simulation discussed in Supplementary Fig. S1, $|E_x| \approx |E_z| >> |E_y|$, and thus $E_r \approx E_z$, which indicates the occurrence of transverse spin angular momentum (T-SAM) [36]: $\sigma^+ = E_r - i\,|E_x|$ in the meridional $zx$-plane (an animation demonstration can be found in Supplementary video S1). At the other interface, since the locations of the isotropic/ anisotropic media in the coordinate system are only reversed along $z$-axis, $E_z$ is also reversed as $E_z'$, resulting in an opposite direction of the transverse spin: $\sigma^- = -E_r - i|E_x|$ which indicates that the spin direction is associated with the constitution of the materials' interface. Such a T-SAM featured state suggests the potential for manipulating the spin of the light. Consequently, the electric field vectors in the coupled surface wave mode forms two inversely symmetrical cycloidal trajectories, as shown in Fig. 1(a) and Supplementary video S1. It is not only the electric field, but also the magnetic field (see Supplementary video S2 and S3), that is spinning transversely according to a reciprocal theory [17], which is in stark contrast to the TM mode with only transversely spinning electric field found in the surface plasmon polaritons [37] (SPPs) (animation demonstrations can be found in Supplementary video S4 and S5). Moreover, compared with SPPs, the propagation of such a mode remains directionally controlled by the orientation of the hyperbolic metamaterial, with a concomitant and significant decrease in propagation loss.

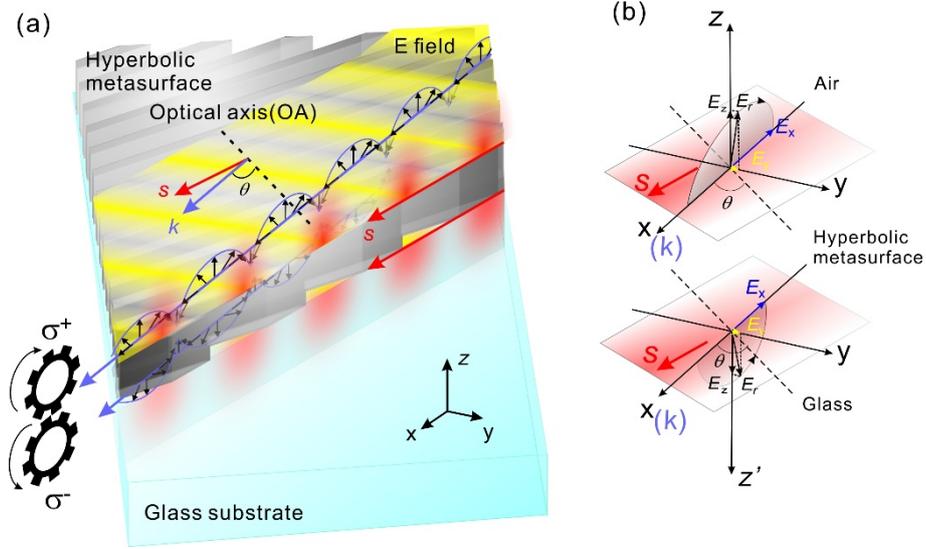

FIG.1. The schematic of the DSW mode. (a) The hyperbolic metasurface is composed of silver grooves on a glass substrate. The DSW mode is confined at the two interfaces between the air/ hyperbolic metasurface and the hyperbolic metasurface/ glass. The red arrow and the purple arrow indicate the Poynting vector $S$ and the wave vector $k$, respectively. Due to the anisotropy, $S$ is not parallel with $k$ that has an angle of $\theta$ with respect to the OA. (b) According to $|E_x| \approx |E_z| >> |E_y|$, $E_r = E_y + E_z \approx E_z$, and because of the $\pi/2$ phase difference between $E_r$ and $E_x$, the electric field shows a transverse spin property at the top interface: $\sigma^+ = E_r - i\,|E_x|$. At the other interface, since the location of the hyperbolic metasurface and the isotropic material is reversed along $z$-axis, $E_z$ is flipped to $-E_z$, then the spin of the electric field is also reversed: $\sigma^- = -E_r - i\,|E_x|$. Therefore, the transverse spinning of the electric fields in the mode is analogous to two rotating gears engaged together, producing two cycloidal traces with mirrored symmetry as shown in (a).

### III. DSW mode

Our experimental demonstration of such a DSW mode is achieved using a hyperbolic metasurface made of silver nano-grooves on a quartz glass substrate, as shown in Fig. 2. The grooved structure is a thin film formed from a silver/ air vertical multilayer structure with its OA lying along the substrate.

Thus, orientation of the hyperbolic anisotropy is defined by the angle $\theta$ between the OA and $x$-axis, as illustrated in Fig. 2(a). The permittivity normal to the grooves is the extraordinary permittivity $\varepsilon_\perp = \varepsilon_e$, and the permittivity along the grooves is the ordinary permittivity $\varepsilon_\parallel = \varepsilon_o$. By using Maxwell-Garnett theory, the anisotropic permittivity of the multilayered structure can be determined by:

$$\varepsilon_o = \varepsilon_m f + \varepsilon_d \left( 1 - f \right) \tag{2a}$$

$$\varepsilon_e = \frac{\varepsilon_d \varepsilon_m}{\varepsilon_d f + \varepsilon_m \left( 1 - f \right)} \tag{2b}$$

where $f$ is the filling ratio of the silver. $\varepsilon_d$ and $\varepsilon_m$ are the dielectric constants of the air and silver respectively. Just from the geometry structure, one may notice that the silver grooves we used here is quite similar to the structure in the work by High et al. [11], which reported a SPP based hyperbolic metasurface according to the waveguide coupling theory. Here, we would like to emphasize that in current design, the silver grooves together with the air form a thin hyperbolic metamaterial film with an indefinite dielectric tensor on a glass substrate. The DSW modes contains two DSWs at the two surfaces of the metamaterial film while coupled together under the small thickness. The silver grooves in Ref. 11 are located on a silver film and thus that structure actually has only one ridged metal surface to support the SPP. Each groove is working as a waveguide and all the intriguing behaviors are originated from the coupling effect between these waveguides. Further discussion about the DSW mode characteristics rather than SPP will be introduced in the rest of the paper and the supplementary materials. Since the dielectric anisotropy is based on effective medium theory, it does not demand a high quality of the silver film to be single crystal [11] but just a good surface roughness, which simplifies the fabrication. In the fabrication, a hyperbolic metasurface of the grooved structure with $f$ around 54% is patterned by a focused-ion-beam (FIB) milling through a 60 nm thick silver film on a glass substrate, as shown in Fig. 2(a). From using equation (2), the effective permittivities at 633 nm are found to be $\varepsilon_e = 2.3$, $\varepsilon_o = -9.4 +$ 0.26i, with the permittivity of silver $\varepsilon_m$ = -18.3 + 0.48i [38], and the permittivity of air $\varepsilon_d$ = 1. According to the boundary condition of the DSW with hyperbolic anisotropy [21], the allowed angle band of the DSW can be theoretically estimated to cover angles from 27$^\circ$ to 45.6$^\circ$ at the air/ metasurface interface and 29.2$^\circ$ to 90$^\circ$ at the metasurface/ glass interface, the later one of which finally determines angle band of the DSW mode due to the strong coupling. This allowed angle band of the mode is also an important characteristic to differentiate from the SPP mode in waveguides array. Considering the momentum matching in the input coupling, we designed a series of orientations $\theta$ of the hyperbolic anisotropy, 45$^\circ$, 50$^\circ$ and 60$^\circ$ to verify the existence and the directionality of the DSWs mode inside the metasurface, as shown in Fig. 2(a), (b) and (c) respectively. Finally, input and output coupling gratings are also patterned on the silver film next to the metasurface which may couple in a surface wave with its wave vector along the $x$-axis.

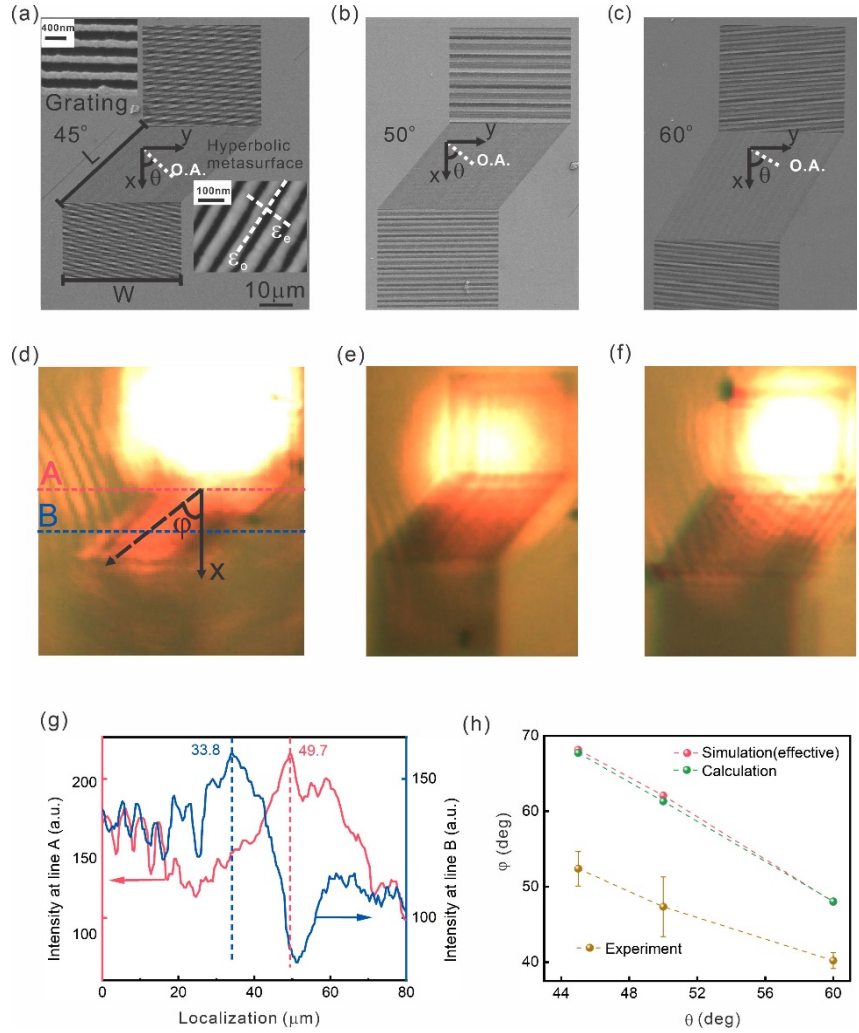

FIG. 2. Characterization of the DSW mode at the metasurface under the three different orientations. (a) An SEM image of the device with $\theta = 45°$. $\theta$ is defined as the orientation angle between the optical axis and $x$-axis. The hyperbolic metasurface consisting of silver grooves with period 80nm and silver linewidth of 43nm is visible. The overall size of the metasurface is designed to be length L= 35µm along the propagation direction with width W = 35µm. Gratings with the same width (W) were fabricated on the silver film next to the metasurface. The period is 300nm with a silver duty cycle of 50%. The scale bar applies to the images (a)- (f). (b) Metasurface with $\theta = 50°$. (c) Metasurface with $\theta = 60°$. (d) Optical image of the DSW propagation in the metasurface. The black dashed arrow indicates the propagation direction $\varphi$ with respect to $x$-axis. Intensity distributions along lines A and B are collected to determine $\varphi$. (e) Optical image of the beam propagation in the metasurface with $\theta = 50°$. (f) Optical image of the beam propagation in the metasurface with $\theta = 60^\circ$. (g) The red and blue curves are the intensity distribution along the line A and line B, respectively. The relative shift between the two peaks which indicates the changing location of the beam is 15.9 μm. (h) The comparison of the propagation direction $\varphi$ between the simulation the experiment and the theoretical calculation. The yellow dots are from the experimental results with an error bar while the pink ones are calculated based on the simulation and the green dots are from the theoretical calculations.

In the experiment, a $y$-polarized laser of 633 nm is focused onto the input grating and then coupled into the hyperbolic metasurface. Light scattered by the inhomogeneities of the metasurface allows us a direct observation at the far field, which clearly shows the beam propagation in all three cases in a CCD camera as evident in Fig. 2(d)- (f). The observation of the surface waves under these three orientations indicates the allowed angle band to be at least 15°. Moreover, after transmitting 35 μm in the hyperbolic metasurface, we can still see very strong coupling at the output gratings, which verifies metasurface's low loss property.

By plotting the intensity distribution across the observed beam, we can determine the propagation direction of the surface wave. As shown in Fig. 2(d), by extracting the intensity distribution along the two lines A and B on the hyperbolic metasurface of $\theta = 45^\circ$, we can track the propagation of the surface wave, indicated by the peaks shown in Fig. 2(g) (details can be found in the Supplementary Fig. S2). The horizontal shift between the two peaks after the surface wave propagation from A to B is 15.9 μm while the vertical distance between line A and line B is 11.9 μm. Thus, the propagation direction is $\varphi = 53.2^\circ$, with respect to the $x$-axis.

Conducting the same analysis on the other two samples where $\theta = 50^\circ$ and $60^\circ$ shows propagation directions of $\varphi = 47.3^\circ$ and $40.2^\circ$ respectively. Numerical simulations and theoretical calculations derived from the transfer matrix method [39-43] are also applied to further validate this behavior. In the simulations shown in Fig. S3, we use a uniform hyperbolic film with the effective permittivities $\varepsilon_e$ and $\varepsilon_o$ but not the actual silver structure in order to avoid the SPP coupling effect between these silver grooves each of which is used as one waveguide. All of these results are summarized in Fig. 2(h). From this analysis, one can see that the value and tendency of the experimental results are in good agreement with the simulation results and theoretical calculations, within tolerable deviations arising from imperfections in the sample's fabrication. Moreover, it should be noted that the beam propagation direction in each case matches identically, being determined by the anisotropy of the hyperbolic metasurface. By using the transfer matrix method, we can calculate the angle dispersion of the wave vector and then plot the Equi-frequency contour of the metasurface, which actually shows a hyperbolic dispersion but not the cosine function in the waveguide coupling theory [11]. This verifies its characterization as a DSW mode rather than a SPP mode which would instead be generated between the silver/air boundary and would propagate along the milled silver grooves.

**IV. PSHE demonstration**

Another intriguing feature of this system is the determination of the transverses spin by the strong coupling between the electric field rotation and the hyperbolic metasurface, causing two different trajectories for the light beams with opposite transverse spins, which is known as the photonic spin Hall effect (PSHE) [44-46]. To characterize PSHE in our system, we rotate the hyperbolic structure by $\theta = 90^\circ$ (OA now along the $y$-axis), so that the allowed DSW mode wave vector may appear on either side of the $x$-axis. As shown in Fig. 3, our structure can support two allowed DSW modes whose wave vectors $\boldsymbol{k}$ are along $\theta$ ($\boldsymbol{k_R}$) with respect to the OA as illustrated by Fig. 3(a) & (c) and $\pi$-$\theta$ ($\boldsymbol{k_L}$) with respect to the OA as illustrated by Fig. 3(b) & (d). Due to the anisotropy of the metasurface, the two modes at the supplementary angles are in mirror symmetry. The numerical simulation in Fig. S5 shows that the two modes maintain the same electric field components in the *x* and *z*-axis, but have opposite directions along the OA, which may flip the spin of the electric fields between the two modes. Therefore, at each interface, DSWs from the two separated modes in Fig. 3(a) and (b) or (c) and (d) are in opposite spins. Furthermore, within each mode, as we discussed above, DSWs at the two interfaces ((a) and (c) or (b) and (d) in Fig. 3) are experiencing opposite spins. Therefore, the PSHE is originated from the interaction between the T-spin of the DSW mode and the anisotropy of the hyperbolic material film while a similar phenomenon occurred to the SPP is actually based on a longitudinal spin [11].

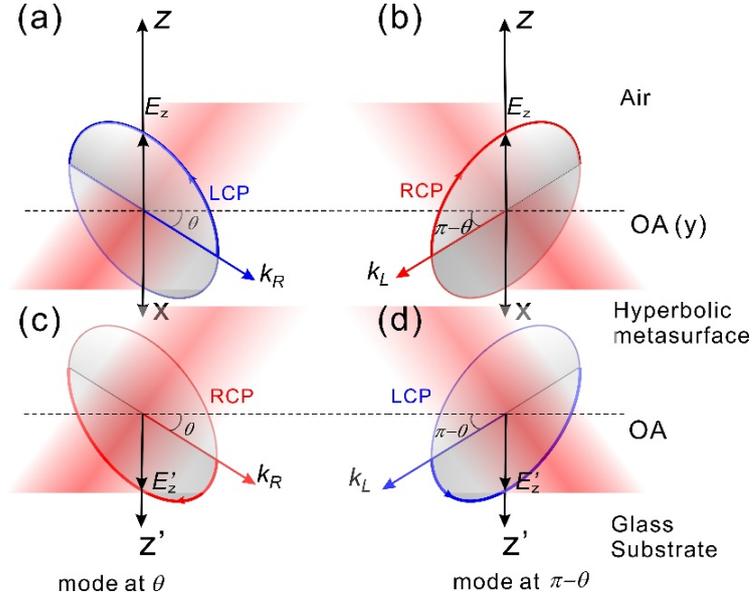

FIG. 3. Schematic of the polarization states at two supplementary angles with respect to the optical axis. On the left side, (a) and (c) are surface waves of the DSW mode with $k_R$, whose angle is $\theta$ with respect to the OA. On the right side, (b) and (d) are the surface waves of the DSW mode with $k_L$, whose angle is $\pi$-$\theta$ with respect to the OA. Between the two interfaces in each mode, spins are reversed due to the flip of the constituent materials. At each interface between the two modes, because of the flip of the electric field along the *y*-axis, the spins of the DSW modes are reversed from $k_L$ to $k_R$ or vice versa.

In the experimental demonstration, the hyperbolic metasurface with the same feature size as above but with its OA along the *y*-axis is fabricated to test the PSHE. A circular hole milled using FIB next to the hyperbolic metasurface is served as the input coupling. In order to enhance the coupling efficiency, an arch grating is also fabricated around the hole to help collect the incident beam and focus it into the hole.

On the other end of the metasurface, an array of silver posts, each having a diameter of 110 nm and being

60 nm in height, is used as the output to couple the beam out while preserving its original polarization features [11], as shown in Fig. 4.

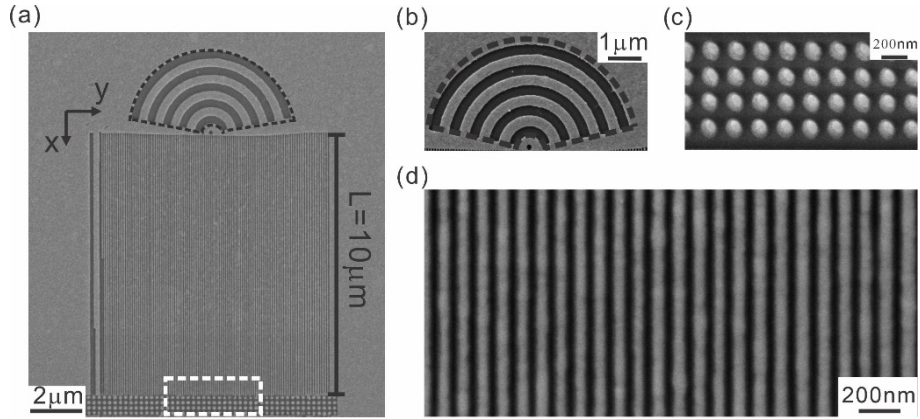

FIG. 4. An SEM image of the PSHE device. (a) The image of the whole structure for the PSHE

demonstration. The metasurface is 10 μm in length. All the features are the same as the samples in Fig.

2. The in-coupling structure is a nano-hole with diameter 140 nm and the arch grating in the black dashed

region is used for enhancing the coupling efficiency. The nano-rod array is used for out-coupling. The white dashed box indicates the out-coupling region used for characterization. (b) Input port with a nano-hole and the arch gratings around it. (c) Output port composed of a nano-rod array. The diameter and the height of the nano-rod are 110 nm and 60 nm, respectively. (d) The magnified image of the hyperbolic metasurface.

In the characterization, firstly a linear polarized beam which is a superposition of the left and right circular polarizations is focused on the input structure, as shown in Fig. 5(a). At the output, there are two bright spots corresponding to two opposite spins in the PSHE theory. To further differentiate the spins of the two out coupling beams, a left circular polarized (LCP) beam is used as the input. Originating from the anisotropy of the hyperbolic metasurface, the Poynting vector $\boldsymbol{S}$ and wave vector $\boldsymbol{k}$ are directed to either side of the $x$-axis (a detailed demonstration can be found in Supplementary Fig. S5, S6 and S7). Consequently, the LCP beam propagates to the right only which preserves the bright patch at the output on the right-hand side, but eliminates the one on the left. By contrast, when a right circular polarized (RCP) beam is used as the input, we observe a reversed result, with a bright spot visible on the left. These different cases are shown in Fig. 5(c) and (e). Figure 5(b), (d) and (f) present the locations of each spot by analyzing the intensity distributions along the center of the out-coupling spots. From these results, the peak positions arising from the left and right circular polarizations in Fig. 5(d) and (f) match the positions of the two spots generated from the linear polarization in Fig. 5(b). We also perform a numerical simulation to show the opposite transverse spins of the two output spots, which confirms the PSHE phenomenon (Supplementary video S6).

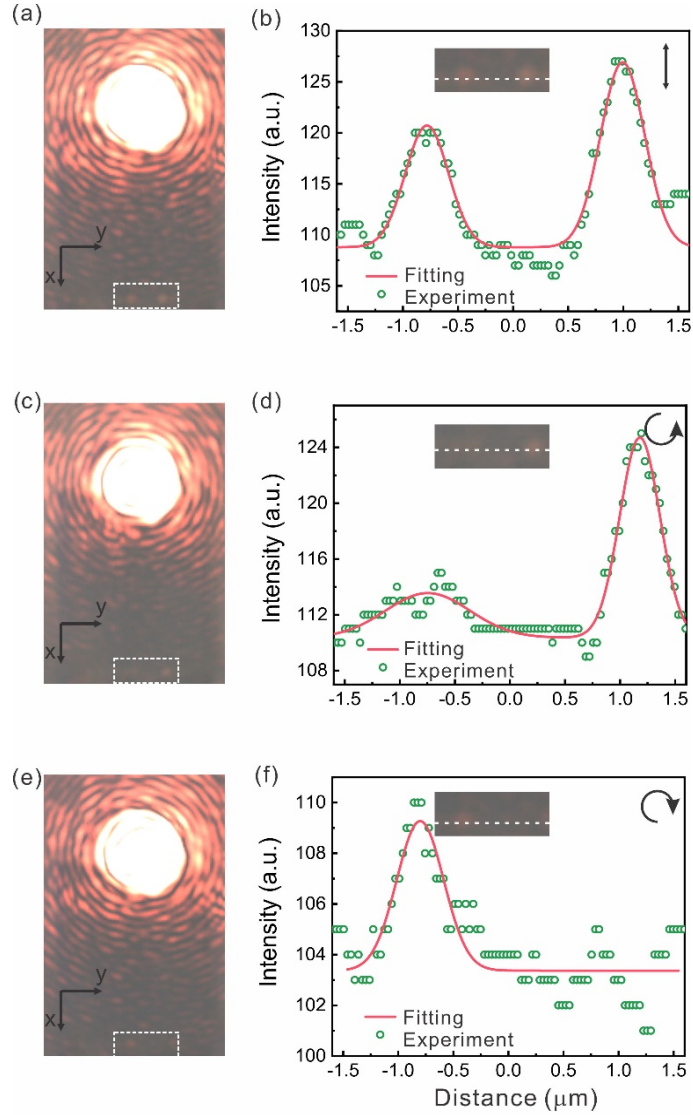

FIG. 5. Experimental demonstration of PSHE based on DSW mode. (a) The image of the PSHE with linear polarization. The incident light at the chosen wavelength of 633 nm is focused onto the input nano hole and propagates along the device in Fig. 5(a). There are two spots at the output structure marked with a white dashed box at exactly the same location in the SEM image in Fig. 4(a). (b) The intensity distribution along the dashed line across the two output spots. (c) The image with left circular polarized incident light. Only one spot is coupled out at the right side of the nano-rod array. (d) The intensity distribution along the dashed line at the output shows an accentuated right peak. (e) The image with right circular polarized incident light. One spot is coupled out at the left side of the nano-rod array. (f) The intensity distribution along the dashed line at the output shows an accentuated left peak.

**V. Conclusions**

We have experimentally demonstrated a DSW mode propagating in a hyperbolic metasurface at a visible frequency. The mode consists of two coupled DSWs interacting between the two interfaces of the hyperbolic metasurface. Regarding the DSW mode itself, its allowed angle band was shown to be at least 15°, significantly larger than those of the conventional DSW systems due to the strong anisotropy of the hyperbolic metasurface. Furthermore, the DSWs measured in our system propagate at least 35 μm, exhibiting a much lower loss comparing to the SPP systems which promises its excellent performance in the 2-dimensional photonic guiding applications. By analyzing the mode at each interface, we showed that the DSW carries a polarization state with a transverse spin property, whose direction is determined by the constituent materials at the interface. Therefore, within the mode, the two DSWs at two interfaces present opposite spins due to the flip of the constituent materials of the interfaces. On the other hand, by analyzing the DSW modes under the orientations of supplementary angles, the two modes show opposite spins. Consequently, a linearly polarized incident field coupled into the hyperbolic metasurface will split into two beams with opposite spins, which is the PSHE. This work may facilitate many planar photonic devices, such as spin-dependent excitation/ detection, two- dimensional photonic circuits, and modulation devices at the subwavelength level.

**ACKNOWLEGEMENTS**

This work was supported by the Basic Science Center Project of NSFC under grant No. 51788104, National Natural Science Foundation of China under Grant No. 11974203, Nos. 12074420, and 52072203.

**Supplementary information for**

**Steering visible Dyakonov surface waves at the hyperbolic metasurface**

Yan Li, Jingbo Sun*, Yongzheng Wen, Ji Zhou*

State Key Laboratory of New Ceramics and Fine Processing, School of Materials Science and

Engineering, Tsinghua University, Beijing 100084, China

Correspondence and requests for materials should be addressed to: jingbosun@tsinghua.edu.cn

zhouji@tsinghua.edu.cn

**Table of Contents:**

**Other supplementary materials for this manuscript include the following:**

**Video S1: transverse spin of DSWs**

**Video S2: electric field of DSWs**

**Video S3: magnetic field of DSWs**

**Video S4: electric field of SPPs**

**Video S5: magnetic field of SPPs**

**Video S6: electric field of PSHE**

**Section 1. Transverse spin of the Dyakonov surface wave mode**

As shown in Fig. S1, the beam is coupled into the metasurface through the grating. To analyze the polarization state, we cut the beam across, where we obtain the vector plot, as shown in the inset on the right in Fig. S1a and the three components of the electric field, as shown in Fig. S1b. By comparing the amplitudes of the of the $E_x$, $E_y$ and $E_z$ (with z = 1 μm and 1.06 μm, where the two interfaces are located), one can easily conclude that $|E_x| \approx |E_z| >> |E_y|$ and $E_r = E_y + E_z \approx E_z$ at the two surfaces of the hyperbolic metasurface, as illustrated by the two lines in Fig. S1b. Here, $E_r$ is the field out of the *xy*-plane. On the other hand, according to the Eq. (1), the fields at the interface can be expressed as:

$$E_x = -iA_{TM}\beta_c, \quad E_y = A_{TE}, \quad E_z = A_{TM}N \tag{S1}$$

where $k_0$ is the wave vector in vacuum, $N$ is the effective refractive index of the DSW mode and $\beta_c = \sqrt{N^2 - n_c^2}$ . $n_c$ is either the refractive index of the air or the glass substrate. When the distance between the two surfaces is decreasing, the two surface waves get coupled together and form one mode. In equation S1, the phase difference between $E_x$ and $E_z$ is π/2. Therefore, the electric field at the top interface shows a transverse spin at the interface: $\sigma^+ = E_r - i |E_x| = N - i\,\beta_c$. Video S1 shows the animation of the transversely spinning electric field. At the other interface, since the materials system is reversed along *z*-direction, the rotation of the spin is also reversed which shows $\sigma^- = -E_r - i|E_x| = -N - i\beta_c$, implying that the spin direction can be determined by the constituent of the materials of the interface. According to the literature [17], the magnetic field has a reciprocal form of the electric field in Eq. (S1):

$$H_x = i\frac{A_{TE}}{\eta_0}\beta_c, \quad H_y = \frac{A_{TM}}{\eta_0}n_c^2, \quad H_z = -\frac{A_{TE}}{\eta_0}N \tag{S2}$$

where there is also a phase difference of π/2 between the longitudinal and the transverse components of the magnetic field. η0 is the intrinsic impedance of the vacuum. Thus, it implies the transverse spin of the magnetic field. Video S3 is the animation demonstration of the transversely spinning magnetic field on the top interface in the simulation.

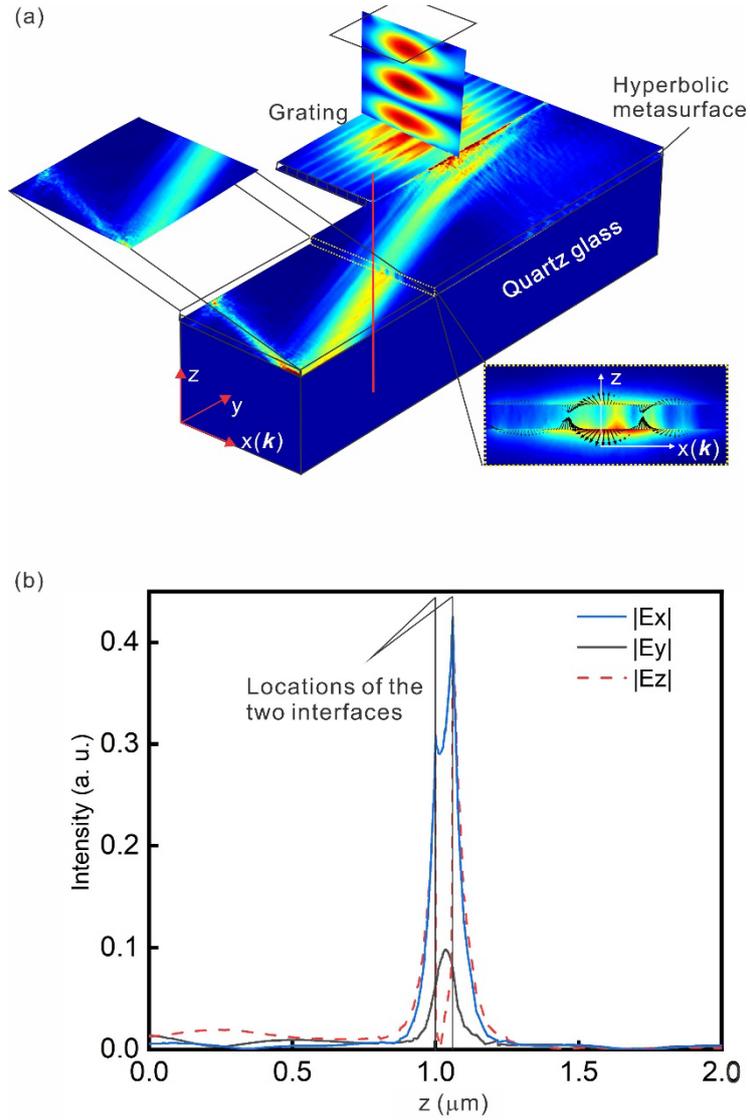

FIG. S1. Numerical simulation of the DSW mode at the two interfaces. (a) Intensity distributions at two interfaces. The inset on the right shows the vectors of the electric field at the cross section cutting along the wave vector $\boldsymbol{k}$. (b) The absolute value of the three components of electric field along the red line in (a).

## Section 2. Propagation direction $\varphi$

In order to avoid confusions of the neighboring peaks in Fig. 2(d), here we plot the intensity distributions on a magnified Fig. 2(d) directly. Due to the scattering, there are several peaks appeared along each line. From Fig. S2, one can easily conclude that the peak with red/blue dash line along A/ B line is the right one which is the actual center of the beam profile. Here we used the coordinates of those two peaks to calculate the propagation direction of the surface wave.

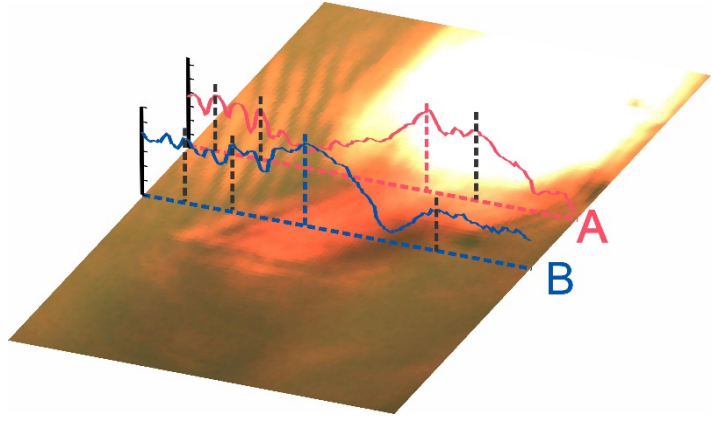

FIG. S2. The propagation process and the intensity distribution of the device in Fig. 2(d).

The intensity distributions along the line A and the line B are presented by the red and blue curves.

The accurate peak positions are marked by the dashed red and blue lines while the scattered peaks around are marked by the dashed black lines.

**Section 3. Simulation results under the different orientation angles $\theta$**

According to the effective medium theory (EMT), the extraordinary and ordinary permittivity of the hyperbolic metasurface is $\varepsilon_e = 2.3$ and $\varepsilon_o = -9.4+0.26i$, respectively. Under a certain orientation angle $\theta$

with respect to x axis, the dielectric tensor of the hyperbolic metasurface can be written as:

$$\begin{bmatrix} \varepsilon_e cos^2\theta + \varepsilon_o sin^2\theta & (\varepsilon_e - \varepsilon_o)cos\theta sin\theta & 0 \\ (\varepsilon_e - \varepsilon_o)cos\theta sin\theta & \varepsilon_o cos^2\theta + \varepsilon_e sin^2\theta & 0 \\ 0 & 0 & \varepsilon_o \end{bmatrix} \quad (S3)$$

Here, we did the numerical simulation on the hyperbolic metasurface structure with $\theta = 45^o$, $50^o$ and

$60^o$. The permittivity of air and glass substrate are 1 and 2.1 respectively. Simulation results of the beam

propagation behavior are shown in Fig. S3, which are used to calculate the propagation direction $\varphi$ in

Fig. 2(h).

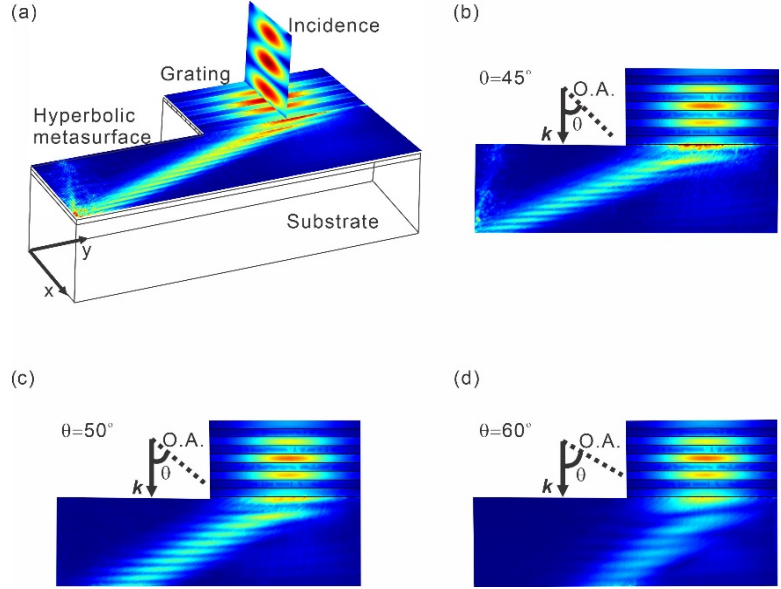

FIG. S3. Simulation results for DSW mode with $\theta = 45°$, 50° and 60° respectively. (a) Perspective view of the simulation result with $\theta = 45°$. Input beam is incident on the input grating and then coupled into the DSW mode in the hyperbolic metasurface. (b) Top view of the simulation with $\theta = 45°$. (c) Top view of the simulation with $\theta = 50°$. (d) Top view of the simulation with $\theta = 60°$. The propagation direction $\varphi$ is determined by examining the intensity distributions along those white dashed lines.

For further demonstration the propagation behavior of DSWs modes, we plot the angle dispersion and Equi-frequency contour (EFC) based on transfer matrix method [39- 43], as shown in Fig. S4. The increasing $\theta$ leads to the decreasing $N$ which may contributes to the decreasing offset eventually. The inset performs the relationship between orientation angle $\theta$ and propagation angle $\varphi$.

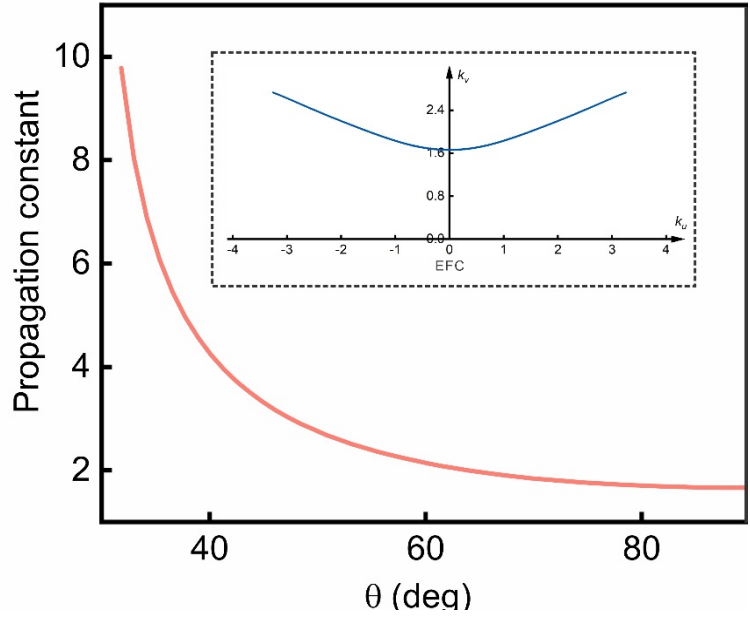

FIG. S4. The angle dispersion of the DSWs mode. The inset shows the EFC of the DSWs modes. The $k_u$ and $k_v$ are the abscissa and ordinate of the EFC while the OA is along $k_u$. The directions of $k$ and $S$ are marked by black and green arrows. The green, pink and blue squares depict the positions under $\theta = 45^\circ$, $50^\circ$ and $60^\circ$, respectively. The propagation angle $\varphi$ equals to the included angle between the Poynting vector (green arrow) and wave vector (black arrow).

## Section 4. The photonic spin Hall effect (PSHE) analysis

In the PSHE demonstration, the optical axis of the hyperbolic metasurface is now oriented along the $y$-axis. Figure S5 shows the field distributions and the intensity distribution at the hyperbolic metasurface/ glass interface with a linearly polarized incidence. After being coupled from the nano hole, the beam splits into two beams. In Fig. S5b, we can label the directions of the wave vectors, which have the angles of $\theta$ and $\pi$-$\theta$ with respect to the optical axis, corresponding to $k_R$ and $k_L$, respectively. By comparing the distributions of the electric field components along the three directions, $E_x$ and $E_z$ are symmetrical with the $x$-axis being the symmetry axis (Fig. S5a, c), but $E_y$ are in opposite directions between the two separated beams (Fig. S5b), which causes the flip of the spin between $k_L$ and $k_R$. We also plot the vector of the $E$ field along the dashed line in Fig. S5d, as shown in Fig.S6. The animation of the electric field is also attached in the Supplementary video S5, which also verifies the opposite spins of the two beams from the linearly polarized incidence.

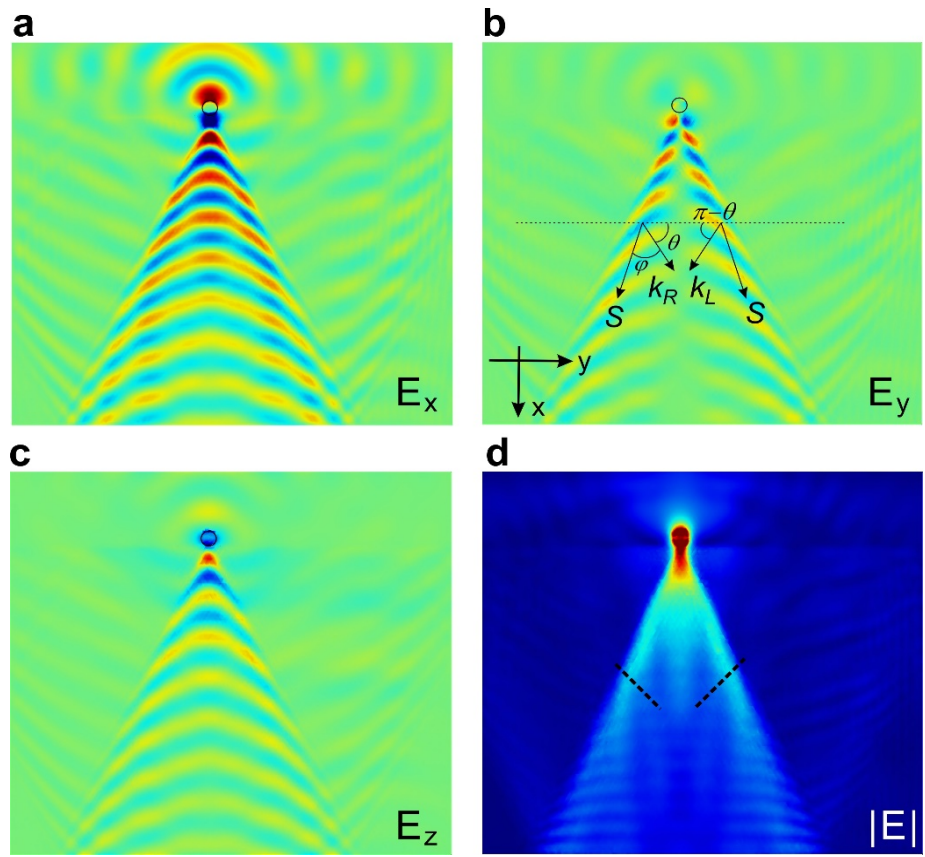

FIG. S5. The PSHE simulation with linearly polarized incidence at the hyperbolic metasurface/ glass

interface. (a) $E_x$ distribution. The coupled Dyakonov surface mode splits into two beams. (b) $E_y$

distribution. (c) $E_z$ distribution. (d) Intensity distribution. The black dashed lines are along the wave vectors of each beam.

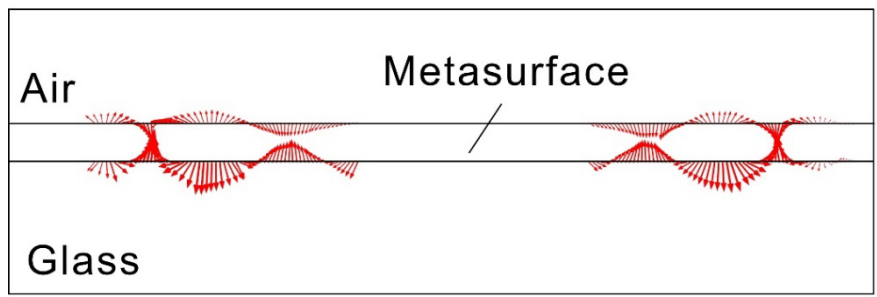

FIG. S6. Transversely spinning E field along the dashed line in Fig. S5d.

The directions of the wave vector $\boldsymbol{k}$ and Poynting vector $\boldsymbol{S}$ are described by the angles of $\theta$ and $\varphi$ with respect to the optical axis and $\boldsymbol{k}$, respectively. Based on the transfer matrix method [41], we can numerically calculate $\varphi$ with $\theta$ ranging from 33º- 90º which is in the allowed angle range of the current hyperbolic metasurface. As shown in Fig. S7, in the entire allowed angle range, $\varphi + \theta$ is larger than 90º, which means the $\boldsymbol{k}$ and $\boldsymbol{S}$ are at the two sides of the $x$-axis. This can also be verified in the numerical simulation as shown in Fig. S5.

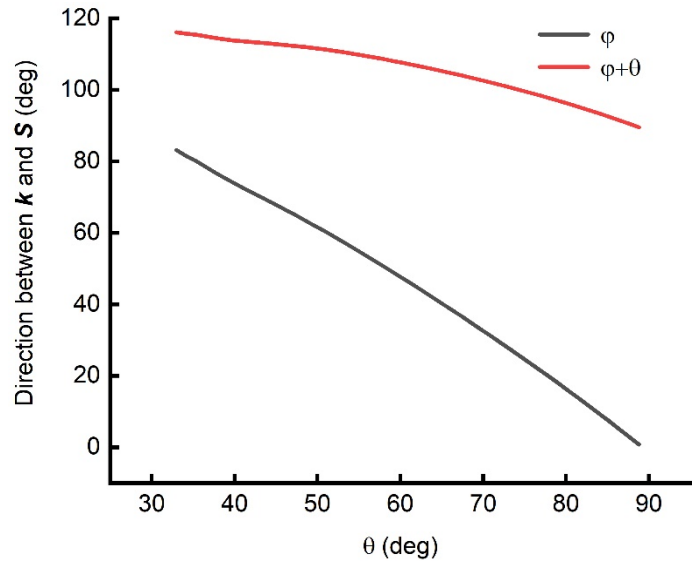

FIG. S7. Relation between the directions of the wave vector $\boldsymbol{k}$ and Poynting vector $\boldsymbol{S}$.

**Section 5. The experimental setup**

For the experimental demonstration of the DSW, a laser at 633 nm (DH-HN250, Daheng Optics) was first transmitted through a linear polarizer and then focused onto the sample with a 20× objective lens (Olympus). The scattered light was collected by the objective and then imaged by a CCD camera. For the investigation of PSHE, to enhance the coupling, a 100× objective lens (Mitutoyo Apochromatic) was used to focus the incident beam onto the input coupling nano-hole. A λ/ 4 wave plate was put after the linear polarizer to generate left/ right circular polarization in turn to test the PSHE (A schematic of the setup is shown in Fig. S8). The white light source was used as the illumination in the imaging system to locate the input grating.

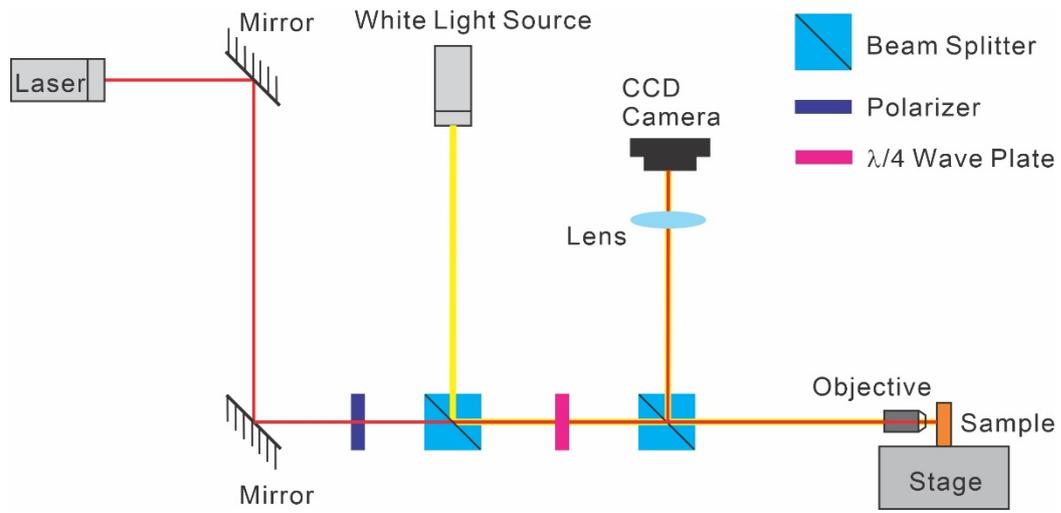

FIG. S8. Experimental setup.